# Chemical Foundation Model Guided Design of High Ionic Conductivity Electrolyte Formulations

*Murtaza Zohair[a,*,†], Vidushi Sharma[a,†], Eduardo A. Soares[b], Khanh Nguyen[a], Maxwell Giammona[a], Linda Sundberg[a], Andy Tek[a], Emilio A. V. Vital[b], and Young-Hye La[a,*]*

[a] IBM Research, San Jose, CA 95120, USA

[b] IBM Research, Rio de Janeiro, Rio de Janeiro 20031-170, Brazil

[†] Equal contribution by authors

[*] Corresponding authors. E-mails: mzohair@ibm.com; yna@us.ibm.com

**ABSTRACT**

Designing optimal formulations is a major challenge in developing electrolytes for the next generation of rechargeable batteries due to the vast combinatorial design space and complex interplay between multiple constituents. Machine learning (ML) offers a powerful tool to uncover underlying chemical design rules and accelerate the process of formulation discovery. In this work, we present an approach to design new formulations that can achieve target performance, using a generalizable chemical foundation model. The chemical foundation model is fine-tuned on an experimental dataset of 13,666 ionic conductivity values curated from the lithium-ion battery literature. The fine-tuned model is used to discover 7 novel high conductivity electrolyte formulations through generative screening, improving the conductivity of LiFSI and LiDFOB based electrolytes by 82% and 172%, respectively. These findings highlight a generalizable workflow that is highly adaptable to the discovery of chemical mixtures with tailored properties to address challenges in energy storage and beyond.

**INTRODUCTION**

Discovering novel chemical mixtures with tailored properties is a critical roadblock to the development of new technologies in drug discovery, environmental remediation, catalysis, cosmetics, and electrolytes for rechargeable batteries[1–5]. The formulation design space grows exponentially with each additional component and interactions between components result in a complex mapping of their chemical and compositional design to a target property[6,7]. As a result, formulation design and optimization is often a time-consuming trial-and-error process.

In the case of battery electrolyte formulations, where research can incur significant materials and testing costs, methods that reduce the number of experiments to reach optimal designs would be extremely valuable[8]. Currently, $LiPF_6$ is the industry standard salt due to its balance of solubility, transport properties, stable solid electrolyte interphase (SEI) formation ability, and chemical stability[9,10]. Trial-and-error testing over the last 30 years has optimized the mixture of linear and non-linear carbonate solvents to pair with $LiPF_6$ in battery electrolytes[11,12]. Several promising alternative salts, such as LiDFOB, offer advantages in terms of thermal stability and SEI stabilization, but suffer from low conductivity in carbonate-based solvents[13]. Insufficient conductivity has been identified to contribute to degradation in commercial battery chemistries via lithium metal plating on the graphite anode[12,14–16]. Therefore, identifying solvent mixtures that enhance ionic conductivity is critical step towards the adoption of new salts. Improving ionic conductivity remains a challenging problem in electrolyte research due to its complex relationship with the molecular geometry of individual constituents in a formulation[17–19]. Addressing this challenge requires efficient searching of the vast and mostly unexplored formulation design space.

The emergence of machine-learning (ML) has ushered in a new paradigm of models that leverage data-driven approaches to learn structure-property relationships for the design of novel battery materials[17–19]. Fewer studies have focused on extending these approaches to electrolyte formulations due to the indirect relationship between formulation property and the properties of constituent molecules[23–31]. One straightforward approach to optimizing electrolyte concentration is to use a surrogate model like Gaussian Process Regression that can guide the exploration of a fixed component system based on the concentration of components[29–31]. These black-box methods have often been paired with high-throughput platforms to guide optimization campaigns faster than design of experiment (DOE) and random searching, but are limited to interpolation within a design space of fixed components. To continue design exploration outside fixed components, using chemical motifs or molecular properties as descriptors is a more favorable approach. However, defining and compiling such descriptors could be difficult in the vast chemical space of battery-relevant molecules[32–34]. This issue can be overcome by using models that can directly map the molecular structure to the target property, thus basing the ML input on the underlying physical chemistry, making the approach generalizable to the vast chemical design space beyond the limitations of available data or descriptors[35,36]. An example would be graph convolution networks (GCNs) that can operate on molecular graphs to encode the molecular topology as an input, which have been widely used to predict properties of molecules, materials, mixtures[23–26], as well as battery performance[27,28]. To scale the GCN for formulations, the graph representations of the molecules in each formulation are aggregated into a formulation descriptor that is updated during the training process. We previously demonstrated the advantages of encoding molecular properties through pre-training the GCN, such as HOMO-LUMO energy levels and electric moment, on the accuracy of predicting formulation performance in small-data regimes common for experimental battery data[27,28]. Though the use of pre-trained molecular structural descriptors (like GCN) in formulation representation overcomes the issue of high degrees of freedom and data limitations, this approach is not efficient due to the reliance on the labeled datasets (HOMO-LUMO and electric moment for molecules) for the pre-training step. The predictive accuracy of these models relies on the relevance of pre-training labels towards the formulation property of interest; therefore, limiting the generalization of the approach. In recent years, transformer-based models trained on large unlabeled text-based representations of molecules (SMILES) have been transformative for the cheminformatics field due to their success in predicting a wide range of molecular properties with high accuracy[37–40]. By utilizing self-supervised learning capabilities, these chemical foundation models reduce the dependence on labeled data and task-specific features[41]. Despite this, the customization and use of these chemical language models for the development of advanced material systems such as formulations remains limited[42,43].

In this paper, we introduce a data-driven model to predict ionic conductivity (IC) of liquid electrolyte formulations, called SMI-TED-IC, that is based on a previously published and open-sourced chemical foundation model SMI-TED (SMILES Transformer Encoder Decoder). We customize a string representation of formulations and fine-tune SMI-TED[44] with a curated dataset of 13,666 electrolyte formulations and their respective ionic conductivity measurements from the lithium-ion battery literature[45–51]. The model is then used to screen through $10^5$ computationally generated electrolytes and identify novel electrolyte formulation designs with improved ionic conductivity that are further validated experimentally. The chemical language model aided formulation discovery workflow is schematically summarized in **Fig. 1a**. The generalizability of foundation models across chemical space enables the exploration of new formulation designs with optimal properties.

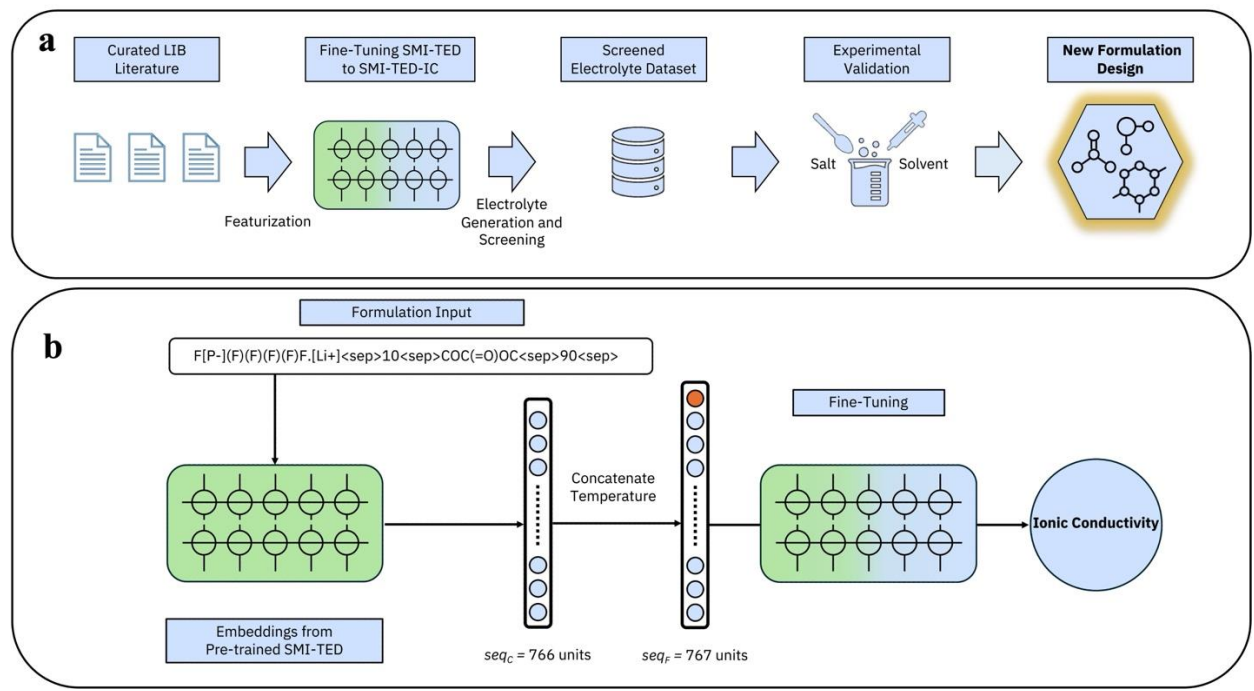

**Fig. 1: Schematic illustration of formulation design workflow and input featurization.** (a) Chemical foundation model (SMI-TED) guided discovery workflow for electrolyte formulations. (b) Schematics describing fine-tuning of SMI-TED model to ionic conductivity predicting SMI-TED-IC. The string input for SMI-TED is adapted for formulation design that includes both SMILES and constituent composition. The model is fine-tuned to predict ionic conductivity after concatenation with the temperature.

## RESULTS

**Analysis of Literature Curated Dataset**

To capture the complex relationship between electrolyte composition and ionic conductivity, we curate a chemically diverse dataset of 13,666 electrolyte formulations and their reported ionic conductivities at different temperatures from the lithium-ion battery literature (**Table S1**). These electrolyte formulations are solutions of charge-carrying lithium salts and aprotic organic solvents that have been investigated for use in various lithium-based chemistries. There are 15 unique salts in the dataset with $LiPF_6$, LiBOB, and $LiBF_4$ being the most prominent ones (**Fig. S1a**). At least one salt is present in every electrolyte and the total salt concentration ranges between 5-15 mol% (**Fig. S1b**). The optimal solvent-salt pairings are dependent upon several physiochemical aspects such as solubility, chemical compatibility, and ion-solvent interactions in the solvation layer[52]. Although it is hard to establish a quantitative trend between these properties and the chemical structure of salts and solvents due to complex interplaying relationships, heuristic design rules based on chemical functional groups have been useful in identifying synergistic combinations of salts and solvent such as the common combination of $LiPF_6$ with carbonate-based solvents[7]. The dataset includes 51 unique solvents with diverse chemical structures to provide a broad spectrum of the chemical design space of lithium-ion battery relevant solvents. For analysis, solvents are grouped into 12 solvent classes based on their chemical similarity and labeled by the primary functional group present in the compound (i.e. carbonates, ethers, etc.). **Fig. 2a** shows a mapping of the chemical structure of the solvents using MACCS keys representations, projected into two dimensions using UMAP[53]. MACCS keys are composed of 166 chemical substructure keys and are a common method for calculating similarity metrics between molecules. Thus, proximity of one molecule to another in the UMAP plot suggests similarity in chemical structures[54]. The cyclic compounds (indicated by open circles in **Fig. 2a, b**) are differentiated from their linear counterparts (filled circles in **Fig. 2a, b**). It is apparent from the UMAP plots that the solvents within a solvent class are clustered together, although cyclic compounds (open circles), are structurally distinct from their linear counterparts (filled circles) for all solvent classes. This clustering is also captured in the UMAP projection of the SMI-TED embeddings of the solvent molecules (**Fig. 2b**), which yields distribution of solvents on the basis of structural information that is embedded in the pre-trained foundation model. The comparison of the two UMAP projections highlights the ability of SMI-TED embeddings to differentiate molecules based on their chemical and structural characteristics. Furthermore, SMI-TED embeddings demonstrate more continuity in the chemical map as evidenced by the convergence of cyclic molecules towards the bottom of the plot in **Fig. 2b.** This is in contrast to the UMAP projection of MACCS keys in **Fig. 2a** where continuity in design space is broken due to the limitations of MACCS keys in being solely focused on individual chemical groups rather than global chemical nature of the molecule. Chemical representations such as the SMI-TED embedding have been demonstrated to have advantages over fingerprints in encoding the influence of semi-local and global molecular structures which will influence many downstream prediction tasks[55].

The distribution of solvent classes in the dataset is shown in **Fig. 2c**. The large majority (77%) of electrolytes include carbonates due to their prevalence in commercial lithium-ion batteries. While the other solvent classes are represented in a minority in the literature dataset, this limited representation does not diminish the scope of the present model due to the large-scale pre-training of SMI-TED on 91 million SMILES samples from PubChem. The wide coverage of the chemical information from the pre-training ensures the generalizability of the model to out-of-distribution formulation designs as demonstrated in the next sections. The concentration of the constituents is also an important determining factor in the ionic conductivity of the formulation. The optimal solvent concentration in a formulation varies by solvent class (**Fig. 2d**). For example, the average concentration of carbonates is high (> 80 mol%) because they are used as the primary solvent in most electrolytes. In contrast, solvents such as esters and fluorinated ethers are mainly used as lower concentrations co-solvents. This difference is accentuated by solubility factor. The solubility of most salts is limited in solvents like fluorinated ethers and therefore their concentration must be reduced to avoid phase separation. The analysis of chemical diversity in the dataset allows us to establish the learned design space of the model.

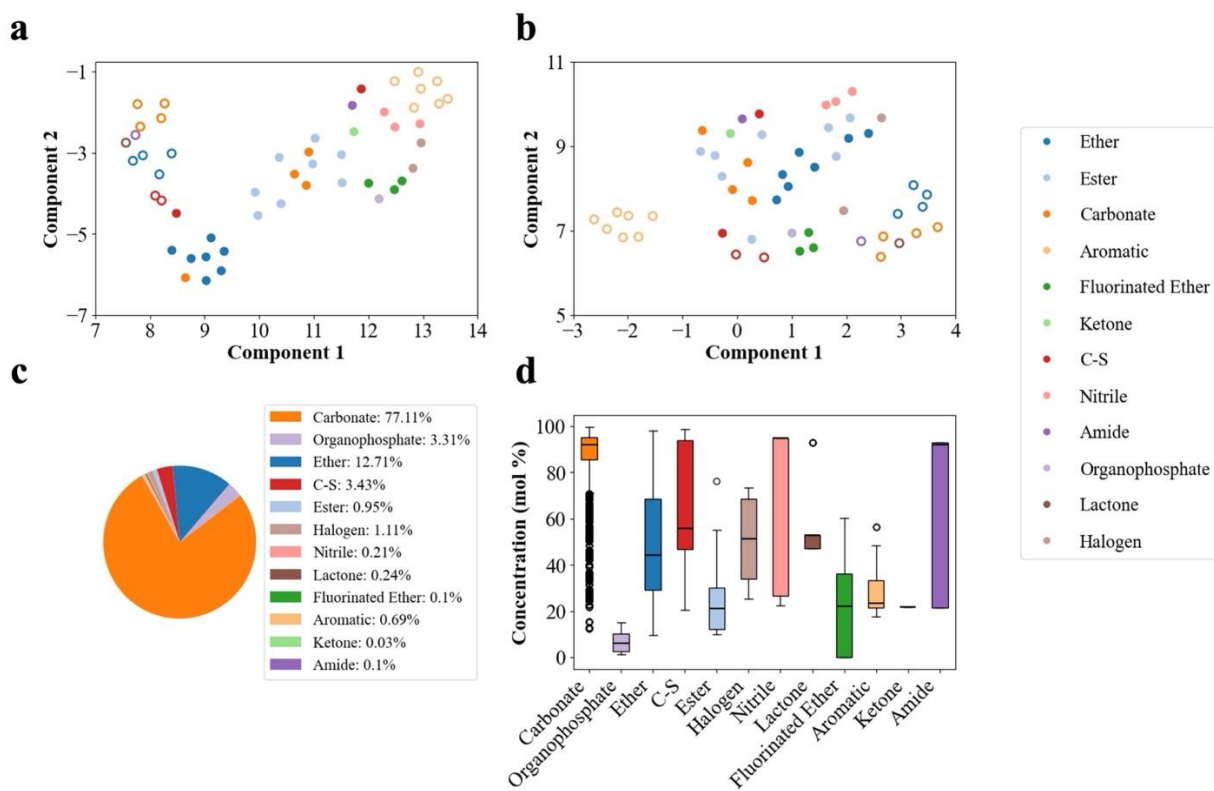

**Fig. 2: Visualization of solvent molecule design space and composition in the dataset.** Two dimensional (2D) UMAP projections of (a) MACCS fingerprints of the solvents in the dataset and (b) SMI-TED embeddings of the solvents in the dataset. The solvents are colored by the assigned solvent classes with open circles corresponding to cyclic compounds and filled circles corresponding to linear compounds. (c) Percentage of electrolyte formulations that contain an instance of a solvent class. (d) The concentration ranges of each solvent class in the dataset.

**Model Performance and Benchmarking**

SMI-TED, a chemical foundation model pre-trained with string representation of 91 million molecules, is fine-tuned with the literature curated electrolyte formulation dataset. The dataset consists of formulations containing up to six electrolyte constituents, where each constituent is represented by its canonical SMILES string. The concentration of each constituent molecule is expressed as the fraction of its mol% in the formulation. The SMILES of molecular constituents in each formulation are combined with their respective compositions into a string representation as demonstrated in **Fig. 1b**, where each component is separated by a <sep> token. This string representation of the formulation and the additional feature temperature is used as an input for SMI-TED fine-tuning (referred to SMI-TED-IC post fine-tuning) while ionic conductivity is used as an outcome label. The final sequence captures the complete chemical, compositional, and conditional (temperature) information of the electrolyte. The complete details of data featurization, augmentation, and regression fine-tuning are described in the **METHODS** section.

The prediction accuracy of the SMI-TED-IC model is shown in **Fig. 3a** as a parity plot. The model is tested on a random 10% subset of the complete dataset. The root mean squared error (RMSE) in ionic conductivity prediction for the test data is 0.1087 $\log_{10}(\sigma)$. The model performance is benchmarked against a formulation graph convolution network model (F-GCN) using the same training and test data splits. The F-GCN model uses GCNs pre-trained with HOMO-LUMO and electric moment to represent the chemical structure of formulation constituents[27,28]. In the present analysis, the SMI-TED-IC model outperforms the F-GCN (**Fig. 3b**) model in conductivity prediction. These results are in agreement with previous reports comparing electrolyte property prediction using chemical foundation models to other machine learning methods[42,43]. Still, the accuracy of the F-GCN model is only slightly lower than the SMI-TED-IC model with a test RMSE of 0.0119 $\log_{10}(\sigma)$. Both SMI-TED-IC and F-GCN map the underlying structural features that influence molecular properties and, therefore, can be used to predict properties in domains with sparse data. To further probe the capability of each model for this task, we evaluate the error on specific solvent classes and salts (**Fig. 3c, d**). Generally, the error among solvent classes and salts is similar for both models, except for fluorinated ether and $LiNO_3$ containing electrolytes, where F-GCN has high prediction errors. These molecules are sparsely represented in the ionic conductivity dataset, with $LiNO_3$ containing only one example in the training set. This finding highlights the crucial difference in the scope of F-GCN and the foundation model, which can be attributed to the differences in their pre-training process. As discussed in the **METHODS** section, the SMI-TED model was pre-trained with over 91 million molecules using a masked language model method, to correctly predict masked molecular tokens based on the rest of the string[38,56]. The success of these chemical foundation models is attributed to the underlying chemical structure rules that the model learns during this self-supervised training process. This conclusion is also supported by a recent study on extrapolation in molecular property prediction, which has demonstrated the advantage of self-supervised pre-training to uncover the relative trends of molecular properties[57]. The F-GCN model on the other hand incorporates a GCN pre-trained on the molecular properties of only 500 lithium-ion battery relevant molecules[27,28]. Thus, the carryover of the supervised pre-training used in the F-GCN model to target electrolyte properties will be contingent upon the relationship between pre-training labels and target formulation properties. Furthermore, the size of labeled datasets will usually be orders of magnitude smaller than unlabeled molecular databases due to the cost associated with calculating labels (properties). Comprehensive mapping of the chemical space enables the SMI-TED-IC model to extrapolate with as few as one example in the training dataset. This demonstrates the unique capability of chemical foundation model-based approaches in exploring the edges of the design space.

A comparison of the literature on machine learning models for liquid electrolyte ionic conductivity prediction are shown in **Fig. 3e**[25,26,30,45]. The number of unique molecules is used as an indicator of chemical diversity and represented by the diameter of the marker. The studies that use models such as Gaussian process regression and linear regression with only concentration as features (dark blue and green markers) are highly accurate, but are constrained to a small number of chemical constituents, and therefore unable to extrapolate to new molecules[30,45]. The work from Zhu and colleagues (purple marker) utilizes a graph and physics-based model that use chemical representations as input, but they focused on optimizing a fixed, five component system. Another graph-based model is applied in the study by Zhang and colleagues to explore a diverse chemical space that included both polymeric and small-molecule containing electrolytes[25]. However, the accuracy across this broader design space is lower (pink marker) than the other models, which can be attributed to the large diversity and small dataset size. By contrast, the F-GCN model used in the present work achieves the second-best accuracy (yellow marker) over a diverse, large dataset, only second to the SMI-TED-IC due to the limitations in extrapolation as discussed above. The work presented here is the first to demonstrate prediction accuracy on-par with the state-of-art models in the literature over a large and chemically diverse dataset. For this reason, the SMI-TED-IC model is next used to screen novel electrolyte designs to identify high performing formulations.

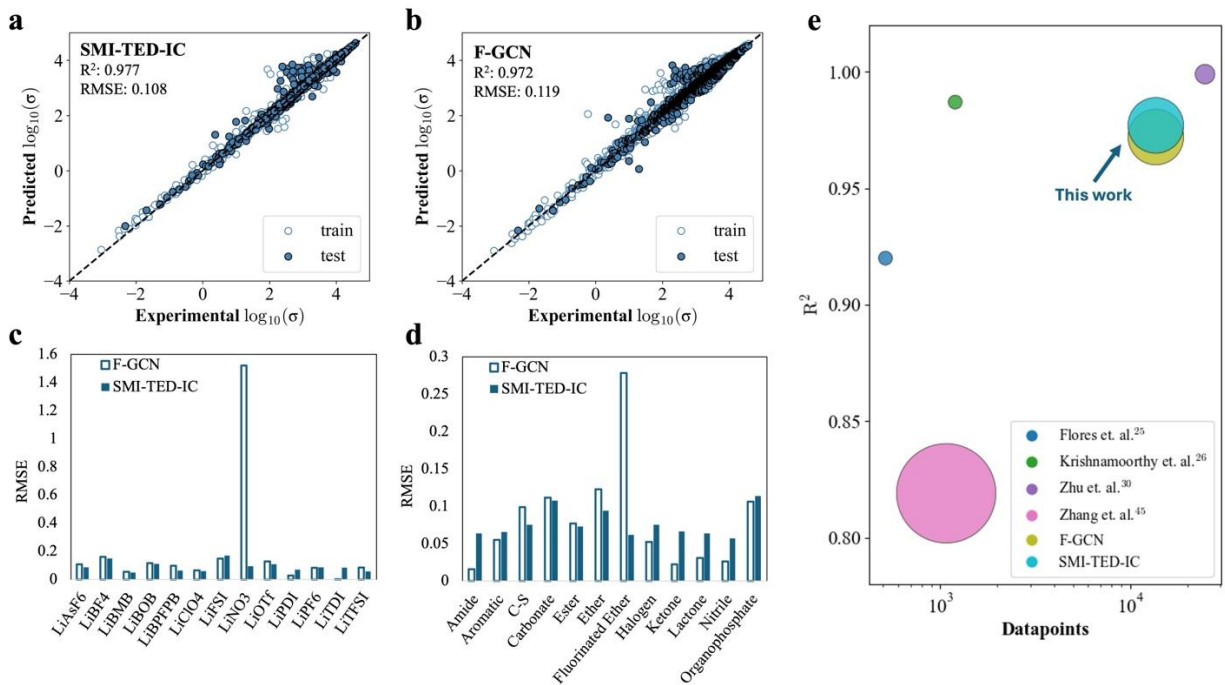

**Fig. 3: Model error benchmarking and comparison to literature.** Parity plots showing predicted ionic conductivity vs. the experimental values in the test dataset for (a) the SMI-TED-IC model and (b) F-GCN model. RMSE of both models on the test dataset for electrolytes containing each unique (c) solvent class and (d) lithium salt. (e) Comparison of predictive accuracy among literature reports of machine-learning models for predicting ionic conductivity literature. The size of the marker corresponds to the chemical diversity in the dataset.

**Design of Novel High Conductivity Electrolytes**

The demonstrated generalizability of the SMI-TED-IC model leads us to probe unexplored areas of the electrolyte design space to discover novel high conductivity formulations. To that end, a "generative screening" campaign is performed by first computationally generating $10^5$ electrolyte formulations that are then screened by the SMI-TED-IC model. To screen for novel high conductivity electrolytes, it must first be noted that the desired ionic conductivity for a lithium-based battery will depend on multiple factors such as cell chemistry and electrode design. Prior electrolyte discovery efforts have chosen various ionic conductivity thresholds for screening novel electrolytes[24,58]. Here we choose a threshold of 10 mS/cm based on the conductivity of 1M LiPF$_6$ in EC/DMC/EMC (1:1:1 v/v/v)[59,60], an electrolyte formulation currently used for evaluating lithium-ion batteries.

The generated dataset samples the immense combinatorial design space of all single salt and solvent combinations from the shortlisted set of molecules available for validation (details in **METHODS** section). Generated electrolytes contain up to five solvents of from any class, resulting in new and more complex electrolyte formulations compared to the literature dataset. The formulations from the generated dataset and room temperature formulations in the literature dataset are visualized through UMAP projections of the SMI-TED-IC formulation embeddings (**Fig. 4a**). Here, the distance between embeddings reflects the relationship between molecular structure and ionic conductivity learned during the fine-tuning process. The separation of the embeddings from the literature and generated dataset in the UMAP plot indicates that the new salt and solvent combinations in the generated dataset are outside the training set distribution. To investigate the finer features of each large cluster, UMAP projections of the individual datasets are generated. The electrolytes in both UMAP plots form distinct clusters based on the constituent lithium salt (**Fig. 4b**, **c**). This is expected given the single-salt constraint on formulation design and because the structural differences between the lithium salts will have a greater impact on the transport properties of electrolytes than the relatively contiguous structural differences between the organic solvents used in lithium-ion batteries[61,52,62]. The distribution of ionic conductivity in these plots (**Fig. 4d**, **e**) can be used to evaluate the trends the model has learned about the salt domains. In **Fig. 4d**, the high ionic conductivity formulations from the literature dataset are found on the left side of the plot, which is also where the clusters with $LiPF_6$, $LiAsF_6$ and LiBOB containing electrolyte are located, whereas the cluster of $LiBF_4$ containing electrolytes tend to have low ionic conductivity. This mapping is consistent with the solvation and ionic mobility characteristics of these salts in organic solvents[52] and with the range of ionic conductivities for each salt in the literature dataset (**Fig. S2**). The 96 electrolytes from the literature dataset above the ionic conductivity threshold of 10 mS/cm are represented in **Fig. 4d** with star shaped markers and are mostly comprised of $LiPF_6$ and $LiAsF_6$ containing electrolytes. The 3351 formulations screened by the SMI-TED-IC model on the generated dataset (**Fig. 4e**) are dispersed more widely across salt domains, suggesting the potential for high-conductivity formulation designs with new salts.

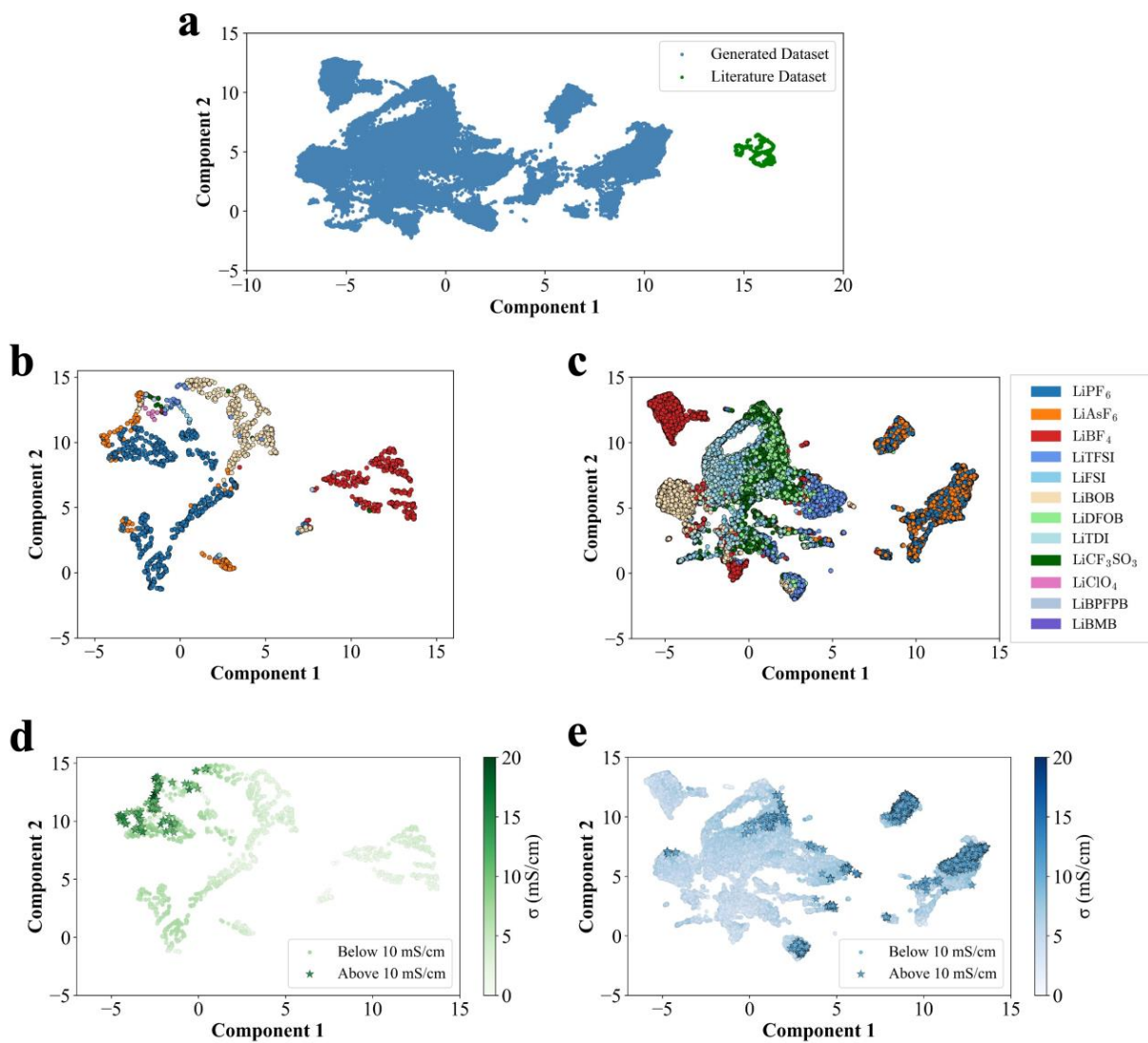

**Fig. 4: Visualization of the SMI-TED-IC formulations embeddings in the generated and literature datasets with UMAP.** (a) UMAP projections of the SMI-TED-IC embedding space to show electrolyte formulation distribution. Datapoints are colored by the source dataset. Individual UMAP projections are colored by the salt present in the electrolyte for (b) the literature dataset and (c) the generated dataset. The same UMAP projections are colored by (d) experimental ionic conductivity in the literature dataset and (e) predicted ionic conductivity in the generated dataset.

To validate the model's ability to identify electrolyte designs with high ionic conductivity, we shortlist 16 electrolyte formulations from the screened group for experimental testing based on the highest predicted conductivity within multiple salt-domains (**Table S3**). Ionic conductivity measurements are made following the procedure described in the supplementary information (SI). **Fig. 5** compares the solvent formulation design and experimental ionic conductivity of the best electrolytes in the literature dataset and the new electrolytes (generated and validated) for three salts. The new electrolyte formulation containing LiPF$_6$ is shown in **Fig. 5a** (generated) and has an ionic conductivity on par with the highest conductivity LiPF$_6$ electrolyte reported in the literature.[46] However, the pie-chart in the figure inset indicates an entirely new solvent formulation design with five different solvent classes to achieve high ionic conductivity. This new design is unique from the literature dataset where electrolytes constitute at most three solvent classes. One of the new components is dimethyl sulfoxide (DMSO), a linear organosulfur (C-S) which does not show up in any of the high conductivity formulations in the literature dataset (**Fig. S3**). Since, the emergent properties of multi-component electrolyte mixtures have been shown to improve ionic conductivity[50,63], discovery of this novel complex formulation emphasizes the scope of chemical foundation models in material discovery, especially within multi-component materials due to the exponential growth of the potential design space with each added component[7].

The ionic conductivity of the new LiFSI electrolyte (generated) is 19.14 mS/cm, which is 82% higher than the best in literature (**Fig. 5b**). Both the literature and generated electrolytes contain dimethyl carbonate. The generated electrolyte additionally contains an ester (methyl formate), a nitrile (propionitrile), and a cyclic ether (2-methyltetrahydrofuran). We note that the model prediction for the conductivity of this electrolyte is 10.52 mS/cm, a significant underprediction. While the model is conservative in the quantitative prediction of ionic conductivity for this underexplored design, the validation of prediction trends suggests that the relationship between chemical design and ionic conductivity is reliably mapped to guide new formulation discovery. Another instance of significant conductivity improvement in a discovered electrolyte is observed with LiDFOB, where the generated electrolyte is 172% more conductive than the literature counterexample (**Fig. 5c**). Interestingly, there are only two formulations with LiDFOB in the training data, both of which have dimethoxyethane (DME) as a sole solvent. While LiDFOB has benefits in SEI formation and thermal stability, the ionic conductivity of LiDFOB electrolytes with ethers and carbonates is lower than the industry standard $LiPF_6$ containing electrolytes[64]. With our model guided approach, we discover an electrolyte with γ-butyrolactone (GBL), diglyme, and 2-methyltetrahydrofuran that improves the conductivity above most electrolytes in the dataset. The synergy between GBL and LiDFOB particularly has led to renewed interest in its use as the primary salt, and recent studies of this system demonstrate promising low temperature battery cycling performance due to high conductivity across a wide range of temperatures[19,65,66]. Recognizing an emerging area of electrolyte design without being trained on it is a key finding that demonstrates the extrapolation capabilities of this model.

Quantitative design rules for formulations are often elusive due to the degree of interactions between variables. The optimal concentration of a component is influenced by its interaction with other constituents in a formulation, and the vast number of possible combinations makes a comprehensive mapping of formulation criteria infeasible[67,68]. Presently, interpretability for chemical representations of formulations faces an additional challenge when using non-linear models, such as the transformer-based SMI-TED, because of their intrinsic dependence on molecular tokens. For this reason, many studies limit their focus to identifying suitable molecular design for individual constituents rather than the complete formulation[69,25]. Our effort to quantify the relationship between the concentration of solvent classes and ionic conductivity in the generated dataset (predicted ionic conductivity) is shown in **Fig. S4** using the Spearman correlation coefficient. Overall, the resultant correlations within the formulation design are mostly weak and non-informative. This stresses the need for the development of interpretability approaches suitable for multivariate design spaces, which would enable new scientific understanding of complex physical systems such as formulations.

Comparison of the ionic conductivity distribution of all validated electrolytes with the literature-extracted dataset in **Fig. 6** attests to the likelihood of finding high performing electrolyte formulations with the present approach. 7 of the 16 validated electrolytes had a conductivity above the threshold value of 10 mS/cm, implying a 44% probability of discovering a new electrolyte design with targeted high performance based on the present approach. The success rate in the inverse design of materials and molecules reported in the literature is highly variable given differences in the materials system, outcome metric, and validation protocols. However, the majority of these studies use first principles calculations to validate the thermodynamic stability of the generated structures rather than experimental validation, reporting a ~10% success rate[70]. While the success rate of 44% from the generative screening approach is promising, it can be further improved by using metaheuristic optimization methods such as genetic algorithms to guide materials generation as shown in a recent study on self-assembling peptide discovery[71]. Such approaches remain to be realized and evaluated for formulation designs in the future works.

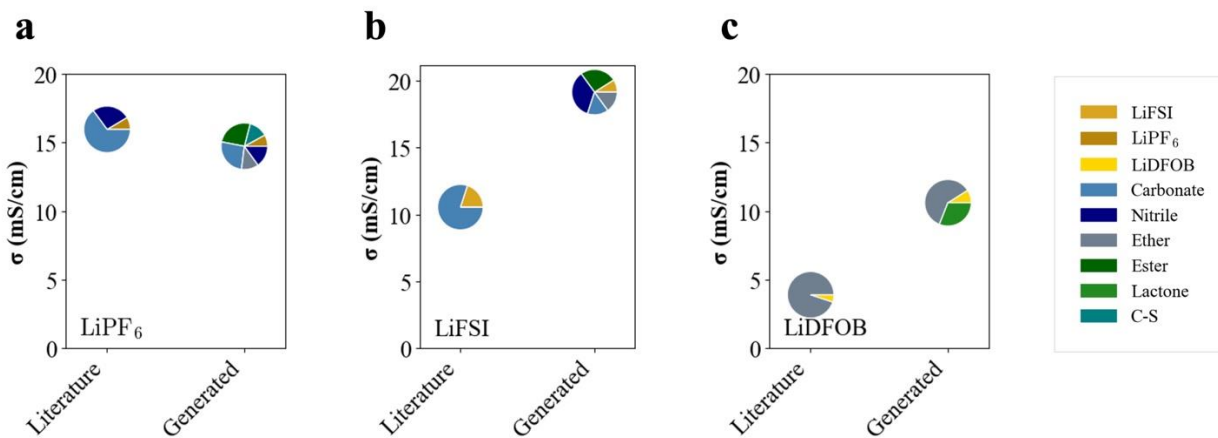

**Fig. 5: Validated novel electrolyte formulation designs for three lithium salts.** Comparison of the most conductive electrolytes from the literature and validated electrolytes from the screened list based on the following salts: (a) LiPF$_6$, (b) LiFSI, and (c) LiDFOB. The concentration of the salt and solvent classes present in each electrolyte are indicated in the inset pie charts.

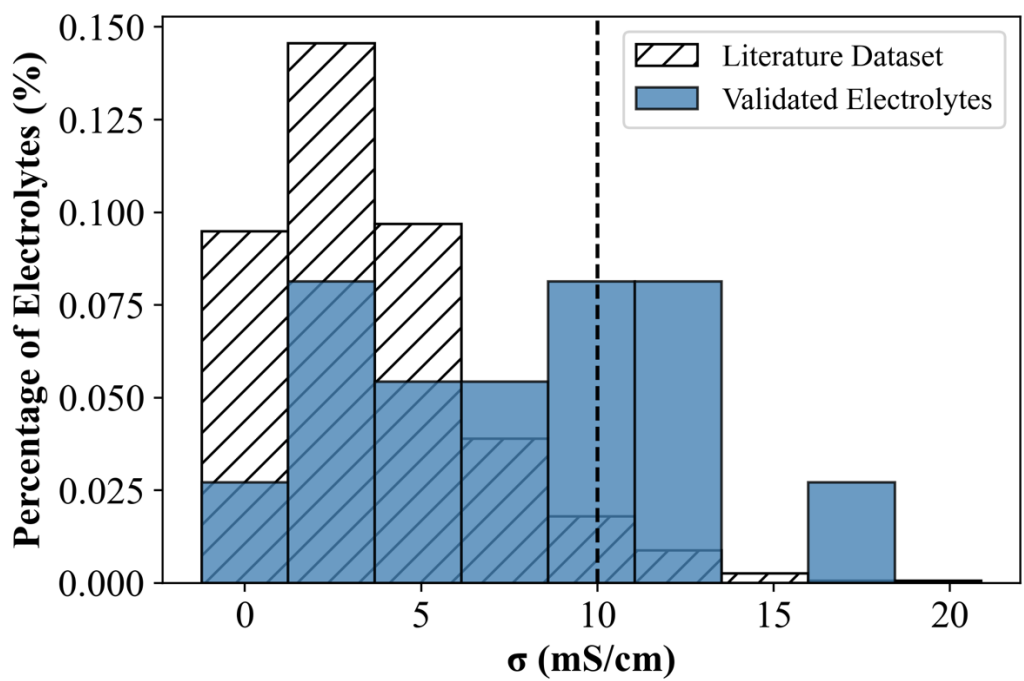

**Fig. 6: Ionic conductivity distribution in the literature dataset compared to foundation model guided designs.** Histogram of experimental ionic conductivity measurements at room temperature from the literature dataset and from novel electrolytes screened by the model and experimentally validated.

**DISCUSSION**

The properties of several materials such as composites and polymer blends are governed by their compositions[58,72]. Accordingly, numerical representations of materials that map their structure and composition to a property have been of significant interest. In the case of battery electrolytes, compositions play a pivotal role in determining the overall performance of the collective mixture. Thus, the incorporation of compositional information in a formulation model is essential. Our approach presents an opportunity to leverage rich chemical embeddings from a pre-trained foundation model and combine them with composition to create a complete representation of multi-constituent materials such as mixtures and formulations. While different modes of aggregation of chemical and composition embeddings can be performed as elaborated in our prior works[42,43], the present framework of formulation representation as a string enables efficient fine-tuning of the language model on the formulation designs, delivering improved generalization. Furthermore, a simplified string representation of formulations/mixtures opens the future scope of applying inverse design strategies for the discovery of such materials.

Integration of a chemical foundation model in the material discovery workflow makes the approach generalizable across a combinatorial chemical space as shown in the present work. The model maintains good prediction accuracy for molecules that are sparsely covered in the training data, like fluorinated ethers and the $LiNO_3$ salt. Furthermore, the model demonstrates a consistent ability to identify high performing formulations across multiple electrolyte chemistries, as validated for three different lithium salts. Lastly, the scope of the SMI-TED-IC model in discovering novel and high performing formulations is highlighted in the discovery of newly reported LiDFOB and LiFSI electrolyte formulations with 172% and 82% improvement in conductivity, respectively, despite training data deficiency. By adopting a pre-trained foundation model for chemical representation, we reduce the dependence on the quantity of labeled formulation datasets, which could be laborious and challenging to acquire in several cases[27,28,34]. Overall, this work showcases the power of foundation models to unlock the full potential of data by bridging knowledge across complex scientific design spaces, paving the way for practical machine learning tools that can actively guide and accelerate experimental discovery.

## METHODS

### Data Curation and Processing

Experimental ionic conductivity values of liquid electrolytes for non-aqueous lithium-based batteries are collected from the lithium-ion battery electrolyte literature (**Table S1**). The criteria for selecting journal articles includes dataset size and chemical diversity. The largest source of data is the CALiSol-23 dataset, which combines electrolyte formulations and ionic conductivity values from 27 journal articles[51]. The electrolyte dataset includes the electrolyte constituents, their respective concentrations, the experimentally measured ionic conductivity, and the temperature that the measurement is taken at. Each molecule in the dataset is represented by the simplified molecular-input line-entry system (SMILES) that are converted to canonical SMILES using the RDkit package to ensure that each molecule has a unique SMILES string. The concentration of each component is converted from the reported unit to molar percentage (mol%) by using the molecular weight and density of the chemical constituent. The temperature is converted from the reported units to Celsius. Additionally, the conductivity values are transformed to a logarithmic scale because ionic conductivity spans multiple orders of magnitude (**Fig. S5**). In the case of a duplicate electrolyte formulation at the same temperature, the ionic conductivity value is averaged.

The temperature values range from 200 to -70 °C (**Fig. S6**). Any datapoints with missing conductivity values are removed from the dataset.

**SMI-TED-IC Model Development**

We employed the SMI-TED a large-scale, open- source model, pre-trained on a curated dataset of 91 million SMILES samples from PubChem, comprising 4 billion molecular tokens, and having 289 million parameters[44]. SMI-TED incorporates an encoder-decoder architecture that significantly enhances model performance and stability over previous foundation models that are based on an encoder-only transformer model and lacked a natural inversion process[42]. The added decoder function in the SMI-TED model overcomes this issue by allowing better representations and denoising. The pre-trained model is customized and fine-tuned for downstream specialized tasks such as predicting an electrolyte formulation property.

A dataset of 13,666 electrolyte formulations and their reported ionic conductivities at different temperatures is curated from the literature. The dataset consists of formulations containing up to six electrolyte constituents, where each constituent is represented by its canonical SMILES string. The concentration of each constituent molecule is expressed as the fraction of its molar% in the formulation. The complete dataset is divided into train, validation, and test subsets by random splitting into 80-10-10 ratios. Next, to generate a good feature representation of electrolyte mixtures, featurization steps are applied sequentially to the train and validation datasets. First, a concatenated sequence (*seq$_A$*) of SMILES of formulation constituents and their respective composition is generated where each component is unified by the <sep> token. As formulant count varies across the whole dataset, formulations containing less than six constituents are padded with dummy features such as SMILES of water ('O'), with their respective composition designated as 0.0. The dummy featurization ensures consistency in the size of input features. Next, the dataset is augmented to include all possible permutations of the constituent SMILES order followed by their respective compositions in the formulations to ensure the model is invariant to the order of formulants in the representation. This step ensures that the model does not learn any unnecessary patterns, such as the sequence of formulation constituents in the dataset, especially since the algorithm is based on a transformer architecture where the learning is sensitive to the order of tokens in the inputs. It is important to highlight that the test dataset is not augmented to preserve fair evaluation of the model. The resultant feature strings (*aug - seq$_A$*) are then transformed into chemical embedding vectors by a pre-trained SMI-TED model (*seq$_C$*) and then aggregated along with their corresponding temperatures at which ionic conductivity has been recorded (*seq$_F$*). The final vector (*seq$_F$*) captures complete chemical, compositional, and conditional (temperature) information of electrolyte formulations and is used as an input feature in the fine-tuning process.

The feature vectors (*seq$_F$*) obtained after the featurization step are used as the inputs in the fine-tuning of SMI-TED model for the downstream task of predicting ionic conductivity. The pre-trained SMI-TED289M model is fine-tuned to learn electrolyte representations (**seq$_F$**) and predict their conductivities. The fine-tuning is conducted over 1200 epochs on the augmented training dataset using a fixed learning rate of 3e-5 and a batch size of 128 *seq$_F$*. During fine-tuning, loss on train and validation datasets are calculated as mean absolute errors (MAEs) and prediction on test dataset is evaluated in root mean squared error (RMSE). A schematic of the fine-tuning process is shown in **Fig. 1b**.

**Electrolyte Candidate Generation and Screening**

To demonstrate the ability of this approach to discover novel electrolyte formulations, we generate 100,000 dummy electrolyte formulations from combinations of the solvents and salts present in the literature extracted dataset and use the SMI-TED-IC model to screen for high ionic conductivity formulations. Certain electrolyte designs would be nonviable as electrolytes such as those consisting of either only solvents or only salts. Also, electrolytes with high salt concentrations (>20 mol%) are screened out to avoid solubility issues. Therefore, generated electrolytes constituting a single salt with a concentration between 5 - 15 mol% are further considered. The number of salts is constrained to one to match the literature dataset, while the solvent concentration range is constrained to above 10%. The molecule pool is filtered for lab availability, and components and concentrations are randomly sampled to generate dummy electrolyte formulations with two to six components. The ionic conductivity of each formulation is predicted at room temperature (20 °C) using the SMI-TED-IC model and a subset of these electrolytes are selected for experimental validation based on ionic conductivity and salt diversity.

**DATA AND CODE AVAILABILITY**

All datasets and codes used in the study will be uploaded in a public GitHub repository.

**REFERENCES**

1. Bannigan, P. et al. Machine learning directed drug formulation development. *Adv. Drug Deliv. Rev.* **175**, 113806 (2021).

2. Sandoval-Pauker, C. et al. Computational Chemistry as Applied in Environmental Research: Opportunities and Challenges. *ACS EST Eng.* **4**, 66–95 (2024).

3. Gomes, G. dos P., Pollice, R. & Aspuru-Guzik, A. Navigating through the Maze of Homogeneous Catalyst Design with Machine Learning. *Trends Chem.* **3**, 96–110 (2021).

4. Xin, H., Virk, A. S., Virk, S. S., Akin-Ige, F. & Amin, S. Applications of artificial intelligence and machine learning on critical materials used in cosmetics and personal care formulation design. *Curr. Opin. Colloid Interface Sci.* **73**, 101847 (2024).

5. Meng, Y. S., Srinivasan, V. & Xu, K. Designing better electrolytes. *Science* **378**, eabq3750 (2022).

6. Narayanan, H. et al. Machine Learning for Biologics: Opportunities for Protein Engineering, Developability, and Formulation. *Trends Pharmacol. Sci.* **42**, 151–165 (2021).

7. Benayad, A. et al. High-Throughput Experimentation and Computational Freeway Lanes for Accelerated Battery Electrolyte and Interface Development Research. *Adv. Energy Mater.* **12**, 2102678 (2022).

8. Ling, C. A review of the recent progress in battery informatics. *Npj Comput. Mater.* **8**, 1–22 (2022).

9. Zhang, S. S., Jow, T. R., Amine, K. & Henriksen, G. L. LiPF6–EC–EMC electrolyte for Li-ion battery. *J. Power Sources* **107**, 18–23 (2002).

10. Aravindan, V., Gnanaraj, J., Madhavi, S. & Liu, H.-K. Lithium-Ion Conducting Electrolyte Salts for Lithium Batteries. *Chem. – Eur. J.* **17**, 14326–14346 (2011).

11. Aurbach, D. et al. Design of electrolyte solutions for Li and Li-ion batteries: a review. *Electrochimica Acta* **50**, 247–254 (2004).

12. Logan, E. R. & Dahn, J. R. Electrolyte Design for Fast-Charging Li-Ion Batteries. *Trends Chem.* **2**, 354–366 (2020).

13. Bian, X. et al. A novel lithium difluoro(oxalate) borate and lithium hexafluoride phosphate dual-salt electrolyte for Li-excess layered cathode material. *J. Alloys Compd.* **736**, 136–142 (2018).

14. Colclasure, A. M. et al. Requirements for Enabling Extreme Fast Charging of High Energy Density Li-Ion Cells while Avoiding Lithium Plating. *J. Electrochem. Soc.* **166**, A1412 (2019).

15. Gallagher, K. G. et al. Optimizing Areal Capacities through Understanding the Limitations of Lithium-Ion Electrodes. *J. Electrochem. Soc.* **163**, A138 (2015).

16. Weiss, M. et al. Fast Charging of Lithium-Ion Batteries: A Review of Materials Aspects. *Adv. Energy Mater.* **11**, 2101126 (2021).

17. Xu, K. Electrolytes and Interphases in Li-Ion Batteries and Beyond. *Chem. Rev.* **114**, 11503–11618 (2014).

18. Seo, D. M. et al. Role of Mixed Solvation and Ion Pairing in the Solution Structure of Lithium Ion Battery Electrolytes. *J. Phys. Chem. C* **119**, 14038–14046 (2015).

19. Hubble, D. et al. Liquid electrolyte development for low-temperature lithium-ion batteries. *Energy Environ. Sci.* **15**, 550–578 (2022).

20. Zhang, H., Wang, Z., Cai, J., Wu, S. & Li, J. Machine-Learning-Enabled Tricks of the Trade for Rapid Host Material Discovery in Li–S Battery. *ACS Appl. Mater. Interfaces* **13**, 53388–53397 (2021).

21. Li, S. & Barnard, A. S. Inverse Design of MXenes for High-Capacity Energy Storage Materials Using Multi-Target Machine Learning. *Chem. Mater.* **34**, 4964–4974 (2022).

22. Merchant, A. et al. Scaling deep learning for materials discovery. *Nature* **624**, 80–85 (2023).

23. Rittig, J. G., Ben Hicham, K., Schweidtmann, A. M., Dahmen, M. & Mitsos, A. Graph neural networks for temperature-dependent activity coefficient prediction of solutes in ionic liquids. *Comput. Chem. Eng.* **171**, 108153 (2023).

24. Kumar, R., Vu, M. C., Ma, P. & Amanchukwu, C. Electrolytomics: A unified big data approach for electrolyte design and discovery. Preprint at https://doi.org/10.26434/chemrxiv-2024-vqtc7 (2024).

25. Zhang, H. et al. Learning Molecular Mixture Property Using Chemistry-Aware Graph Neural Network. *PRX Energy* **3**, 023006 (2024).

26. Zhu, S. et al. Differentiable modeling and optimization of non-aqueous Li-based battery electrolyte solutions using geometric deep learning. *Nat. Commun.* **15**, 8649 (2024).

27. Sharma, V. et al. Formulation Graphs for Mapping Structure-Composition of Battery Electrolytes to Device Performance. *J. Chem. Inf. Model.* **63**, 6998–7010 (2023).

28. Sharma, V. et al. Improving Electrolyte Performance for Target Cathode Loading Using Interpretable Data-Driven Approach. Preprint at https://doi.org/10.48550/arXiv.2409.01989 (2024).

29. Dave, A. et al. Autonomous Discovery of Battery Electrolytes with Robotic Experimentation and Machine Learning. *Cell Rep. Phys. Sci.* **1**, (2020).

30. Flores, E. et al. Learning the laws of lithium-ion transport in electrolytes using symbolic regression. *Digit. Discov.* **1**, 440–447 (2022).

31. Yan, P. et al. Non-aqueous battery electrolytes: high-throughput experimentation and machine learning-aided optimization of ionic conductivity. *J. Mater. Chem. A* **12**, 19123–19136 (2024).

32. Baskin, I., Epshtein, A. & Ein-Eli, Y. Benchmarking machine learning methods for modeling physical properties of ionic liquids. *J. Mol. Liq.* **351**, 118616 (2022).

33. Bilodeau, C. et al. Machine learning for predicting the viscosity of binary liquid mixtures. *Chem. Eng. J.* **464**, 142454 (2023).

34. Kim, S. C. et al. Data-driven electrolyte design for lithium metal anodes. *Proc. Natl. Acad. Sci.* **120**, e2214357120 (2023).

35. Wigh, D. S., Goodman, J. M. & Lapkin, A. A. A review of molecular representation in the age of machine learning. *WIREs Comput. Mol. Sci.* **12**, e1603 (2022).

36. Haghighatlari, M. et al. Learning to Make Chemical Predictions: The Interplay of Feature Representation, Data, and Machine Learning Methods. *Chem* **6**, 1527–1542 (2020).

37. Wang, S., Guo, Y., Wang, Y., Sun, H. & Huang, J. SMILES-BERT: Large Scale Unsupervised Pre-Training for Molecular Property Prediction. in *Proceedings of the 10th ACM International Conference on Bioinformatics, Computational Biology and Health Informatics* 429–436 (ACM, Niagara Falls NY USA, 2019). doi:10.1145/3307339.3342186.

38. Ross, J. et al. Large-scale chemical language representations capture molecular structure and properties. *Nat. Mach. Intell.* **4**, 1256–1264 (2022).

39. Yüksel, A., Ulusoy, E., Ünlü, A. & Doğan, T. SELFormer: molecular representation learning via SELFIES language models. *Mach. Learn. Sci. Technol.* **4**, 025035 (2023).

40. Pyzer-Knapp, E. O. et al. Foundation models for materials discovery – current state and future directions. *Npj Comput. Mater.* **11**, 1–10 (2025).

41. Flam-Shepherd, D., Zhu, K. & Aspuru-Guzik, A. Language models can learn complex molecular distributions. *Nat. Commun.* **13**, 3293 (2022).

42. Soares, E., Sharma, V., Brazil, E. V., Na, Y.-H. & Cerqueira, R. Capturing Formulation Design of Battery Electrolytes with Chemical Large Language Model. Preprint at https://doi.org/10.21203/rs.3.rs-3593035/v1 (2024).

43. Priyadarsini, I. et al. Improving Performance Prediction of Electrolyte Formulations with Transformer-based Molecular Representation Model. Preprint at https://doi.org/10.48550/arXiv.2406.19792 (2024).

44. Soares, E. et al. A Large Encoder-Decoder Family of Foundation Models For Chemical Language. Preprint at https://doi.org/10.48550/arXiv.2407.20267 (2024).

45. Narayanan Krishnamoorthy, A. et al. Data-Driven Analysis of High-Throughput Experiments on Liquid Battery Electrolyte Formulations: Unraveling the Impact of Composition on Conductivity. *Chemistry–Methods* **2**, e202200008 (2022).

46. Hall, D. S. et al. Exploring Classes of Co-Solvents for Fast-Charging Lithium-Ion Cells. *J. Electrochem. Soc.* **165**, A2365 (2018).

47. Hossain, M. J. et al. The Relationship between Ionic Conductivity and Solvation Structures of Localized High-Concentration Fluorinated Electrolytes for Lithium-Ion Batteries. *J. Phys. Chem. Lett.* **14**, 7718–7731 (2023).

48. Watanabe, Y., Ugata, Y., Ueno, K., Watanabe, M. & Dokko, K. Does Li-ion transport occur rapidly in localized high-concentration electrolytes? *Phys. Chem. Chem. Phys.* **25**, 3092–3099 (2023).

49. Bergstrom, H. K. & McCloskey, B. D. Ion Transport in (Localized) High Concentration Electrolytes for Li-Based Batteries. *ACS Energy Lett.* **9**, 373–380 (2024).

50. Wang, Q. et al. High entropy liquid electrolytes for lithium batteries. *Nat. Commun.* **14**, 440 (2023).

51. de Blasio, P., Elsborg, J., Vegge, T., Flores, E. & Bhowmik, A. CALiSol-23: Experimental electrolyte conductivity data for various Li-salts and solvent combinations. *Sci. Data* **11**, 750 (2024).

52. Mauger, A., Julien, C. M., Paolella, A., Armand, M. & Zaghib, K. A comprehensive review of lithium salts and beyond for rechargeable batteries: Progress and perspectives. *Mater. Sci. Eng. R Rep.* **134**, 1–21 (2018).

53. Becht, E. et al. Dimensionality reduction for visualizing single-cell data using UMAP. *Nat. Biotechnol.* **37**, 38–44 (2019).

54. Durant, J. L., Leland, B. A., Henry, D. R. & Nourse, J. G. Reoptimization of MDL Keys for Use in Drug Discovery. *J. Chem. Inf. Comput. Sci.* **42**, 1273–1280 (2002).

55. Qu, J. et al. Leveraging language representation for materials exploration and discovery. *Npj Comput. Mater.* **10**, 1–14 (2024).

56. Devlin, J., Chang, M.-W., Lee, K. & Toutanova, K. BERT: Pre-training of Deep Bidirectional Transformers for Language Understanding. Preprint at https://doi.org/10.48550/arXiv.1810.04805 (2019).

57. Takashige, S., Hanai, M., Suzumura, T., Wang, L. & Taura, K. Is Self-Supervised Pretraining Good for Extrapolation in Molecular Property Prediction? Preprint at https://doi.org/10.48550/arXiv.2308.08129 (2023).

58. Li, K., Wang, J., Song, Y. & Wang, Y. Machine learning-guided discovery of ionic polymer electrolytes for lithium metal batteries. *Nat. Commun.* **14**, 2789 (2023).

59. Xiao, L. F., Cao, Y. L., Ai, X. P. & Yang, H. X. Optimization of EC-based multi-solvent electrolytes for low temperature applications of lithium-ion batteries. *Electrochimica Acta* **49**, 4857–4863 (2004).

60. Zhou, H., Fang, Z. & Li, J. LiPF6 and lithium difluoro(oxalato)borate/ethylene carbonate + dimethyl carbonate + ethyl(methyl)carbonate electrolyte for Li4Ti5O12 anode. *J. Power Sources* **230**, 148–154 (2013).

61. Seo, D. M. et al. Electrolyte Solvation and Ionic Association. *J. Electrochem. Soc.* **159**, A553 (2012).

62. Xie, J. & Lu, Y.-C. Designing Nonflammable Liquid Electrolytes for Safe Li-Ion Batteries. *Adv. Mater.* **37**, 2312451 (2025).

63. Kim, S. C. et al. High-entropy electrolytes for practical lithium metal batteries. *Nat. Energy* **8**, 814–826 (2023).

64. Schedlbauer, T. et al. Lithium difluoro(oxalato)borate: A promising salt for lithium metal based secondary batteries? *Electrochimica Acta* **92**, 102–107 (2013).

65. Lazar, M. L. & Lucht, B. L. Carbonate Free Electrolyte for Lithium Ion Batteries Containing γ-Butyrolactone and Methyl Butyrate. *J. Electrochem. Soc.* **162**, A928 (2015).

66. Gu, Y., Fang, S., Yang, L. & Hirano, S. A safe electrolyte for high-performance lithium-ion batteries containing lithium difluoro(oxalato)borate, gamma-butyrolactone and non-flammable hydrofluoroether. *Electrochimica Acta* **394**, 139120 (2021).

67. Ni, L., Xu, G., Li, C. & Cui, G. Electrolyte formulation strategies for potassium-based batteries. *Exploration* **2**, 20210239 (2022).

68. Yang, Y., Yang, W., Yang, H. & Zhou, H. Electrolyte design principles for low-temperature lithium-ion batteries. *eScience* **3**, 100170 (2023).

69. Bradford, G. et al. Chemistry-Informed Machine Learning for Polymer Electrolyte Discovery. *ACS Cent. Sci.* **9**, 206–216 (2023).

70. Kim, S., Noh, J., Gu, G. H., Aspuru-Guzik, A. & Jung, Y. Generative Adversarial Networks for Crystal Structure Prediction. *ACS Cent. Sci.* **6**, 1412–1420 (2020).

71. Njirjak, M. et al. Reshaping the discovery of self-assembling peptides with generative AI guided by hybrid deep learning. *Nat. Mach. Intell.* **6**, 1487–1500 (2024).

72. Milad, A. et al. Development of ensemble machine learning approaches for designing fiber-reinforced polymer composite strain prediction model. *Eng. Comput.* **38**, 3625–3637 (2022).